\documentclass[journal]{IEEEtran}

\usepackage{graphicx}
\usepackage{amsmath,amssymb,amsthm}
\usepackage{algorithm}
\usepackage{algpseudocode}
\usepackage{cite}
\usepackage{hyperref}

\newcommand{\cC}{\mathcal{C}}
\newcommand{\cD}{\mathcal{D}}
\newcommand{\cE}{\mathcal{E}}

\newcommand{\cI}{\mathcal{I}}

\newcommand{\cO}{\mathcal{O}}

\newcommand{\cT}{\mathcal{T}}
\newcommand{\cU}{\mathcal{U}}

\newcommand{\tr}{\text{Tr}}

\newcommand{\poly}{\text{poly}}

\newtheorem*{theorem5}{Theorem 5}
\newtheorem*{lemma10}{Lemma 10}
\newtheorem*{lemma13}{Lemma 13}
\newtheorem*{lemma9}{Lemma 9}
\newtheorem*{theorem10}{Theorem 10}

\begin{document}

\title{Bootstrapping shallow circuits}

\author{Ning Bao,
\thanks{N. Bao is with the Computational Science Initiative, Brookhaven National Laboratory, Upton, NY 11973, USA, and also with the Department of Physics, Northeastern University, Boston, MA 02115, USA. Email: ningbao75@gmail.com.}
Gün Süer
\thanks{G. Süer is with the Department of Physics, Northeastern University, Boston, MA 02115, USA. Email: gunsuer06@gmail.com.}}

\maketitle

\begin{abstract}
Recently, a polynomial time classical algorithm has been found for learning the shallow representation of a unitary $U$ acting on $n$-qubits, by learning local inversions and then sewing them back together with ancilla qubits and SWAP gates. In this work, we bootstrap local inversion learning (LIL) to optimize quantum circuit depth by learning shallow representations for its sub-unitaries. We recursively cut circuits and apply the LIL algorithm to replace sub-circuits with their shallow representations, if it can be found by the algorithm. If not, we keep cutting until the optimization terminates, either by finding shallow representations or by reaching constant-depth sub-circuits. By replacing sub-circuits with their shallow representations, we hope to obtain some compression of the quantum circuit. Due to the binary search structure, the optimization algorithm has time complexity logarithmic in the depth of the original given circuit.
\end{abstract}

\section{Introduction}

Even though the state of the art of quantum computing is steadily improving \cite{Bluvstein_2023}, we are still far away from a scalable fault-tolerant quantum computer. To benchmark novel quantum algorithms in the noisy intermediate scale quantum (NISQ) era \cite{Preskill_2018}, we are led to techniques that can optimize the central resources used by quantum computers (number of qubits, number of gates, total computation time) and reduce the build-up of errors.

Considering the immense developments in machine learning (ML) over the past decade, it is reasonable that ML can be an important tool in the study, design, and optimization of quantum circuits. However, there are two distinct approaches for bridging quantum computation and machine learning. First, one can use classical machine learning to parameterize a quantum circuit $U(\theta)$ and train it with respect to a cost function, which can be for example the trace distance between the circuit that we want to learn $U$ and the parameterized circuit $U(\theta)$. This is the main idea behind variational quantum algorithms \cite{Peruzzo_2014, McClean_2016, Cerezo_2021_2}. Different ML techniques have also been fruitful in the study of quantum circuit synthesis and optimization, including deep reinforcement learning \cite{fösel2021quantum,daimon2023quantum} and learned guidance function methods \cite{bao2023rubiks}. An orthogonal approach is to implement quantum algorithms as subroutines inside machine learning algorithms, to get similar speedups as one does for well-established quantum algorithms \cite{Biamonte_2017}. Even though both approaches are well-established theoretically, they possess a clear difference in terms of practicality. Machine learning is powerful in practice, but its theoretical foundations are not well understood. One can say the exact opposite for quantum computation; it is very appealing theoretically, but experimentally realizing any of the interesting algorithms on a large scale is very challenging. In this regard, using classical machine learning for the study of quantum circuits is better suited for NISQ applications, while quantum-classical hybrid machine learning algorithms are expected to be much more exciting once there are fault-tolerant quantum computers with appropriate amounts of memory and processing power.

Shallow quantum circuits are constant depth $d = \cO(1)$ circuits. Most notably, shallow quantum circuits can solve certain problems that require depth logarithmic in the number of inputs for classical circuits, hence they are more powerful than their classical counterparts \cite{Bravyi_2018,watts2024unconditional}. However, their power over shallow classical circuits is precisely what makes it hard to learn shallow quantum circuits, since they can generate classically-hard non-local correlations \cite{Bermejo_Vega_2018,Hangleiter_2023,Marrero:2021xmi}. The learning landscape of deep parameterized quantum circuits suffers from the existence of barren plateaus, exponentially large regions in the parameter space that have vanishing gradients, which makes classical learning approaches very challenging \cite{McClean_2018,Holmes_2021,Cerezo_2021_3,Holmes_2022}. While the landscape of shallow quantum neural networks is free from barren plateaus, it is swamped with exponentially many sub-optimal local minima \cite{Anschuetz_2022}. This renders standard local optimization techniques such as gradient descent futile in trying to learn shallow quantum neural networks.

The problem of learning shallow quantum neural networks has recently been overcome in \cite{huang2024learning} by a clever reconstruction of shallow quantum circuits. The authors show that a shallow quantum circuit can be represented using its local inversions, sewn together with ancilla qubits and SWAP gates. This reduces the problem of learning shallow quantum circuits to learning its local inversions. It is proven that polynomial time classical algorithms that use randomized measurement data exist for learning the inversions, and a polynomial time verification of the learned circuit is presented. Therefore the local inversion learning (LIL) algorithm can produce a learned quantum circuit $\hat{U}$ that is constructed out of local inversions of the unitary $U$, and verify that it simulates the original unitary up to a small diamond distance. 

Inspired by these developments, we present a novel optimization scheme that can reduce the total computation time for a given circuit by recursively cutting the circuit in half and learning shallow representations of the sub-unitaries. Given a circuit $U$, it is highly unlikely that the whole circuit can have a shallow representation. However, unitaries corresponding to certain sub-circuits of the whole circuit can have shallow representations. To this end, we recursively apply the LIL algorithm to the circuit that we want to optimize, by cutting the original circuit in half. We start with the original circuit itself. If the LIL algorithm succeeds, it gives us an $d = \cO(1)$ representation of the circuit. If it fails, we cut the circuit in half, e.g. $U = U' U''$, and apply the LIL to each of the sub-circuits corresponding to the unitaries $U'$ and $U''$. If the LIL algorithm succeeds, we replace the circuits with their shallow representations; if not, we keep cutting the circuits until each sub-circuit has been replaced with its shallow representation. This is either achieved by finding non-trivial shallow representations for each of the sub-circuits, or by cutting until we reach single gate unitaries, in which case they are automatically shallow. Due to the binary search structure of the recursive optimization algorithm, the total time complexity of the optimization algorithm is $t = \cO(\log d)$. However, since the LIL for each sub-circuit of the entire circuit is independent from one another, the optimization algorithm is massively parallelizable. This is especially important as one goes lower in the binary tree, since the LIL time complexity remains the same, while the number of sub-circuits increases exponentially by depth.

In this manuscript, we start by giving a brief introduction to local inversion learning of shallow quantum circuits in Section \ref{sec:inversion}. We reiterate how local inversions of a unitary can be used to reconstruct itself and highlight the important features of the local inversion learning algorithm relevant to circuit compression. In Section \ref{sec:compression}, we show how arbitrary quantum circuit depth can be compressed via the iterative application of the local inversion learning algorithm. This is the central result of our work. We end the paper with Section \ref{sec:discussion} by discussing different cut structures, optimization of cut ratios, and further directions.

\section{Learning shallow quantum circuits}
\label{sec:inversion}

\begin{figure}[h]
    \includegraphics[width=\linewidth]{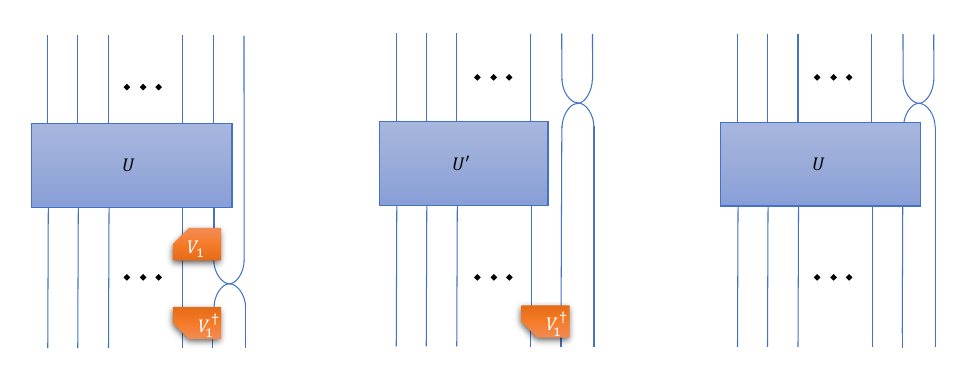}
    \caption{Reconstruction of $U$ via local inversions. For simplicity, we only demonstrate the trick for a single qubit.}
    \label{fig:local-inversion}
\end{figure}

We give a brief overview of the shallow quantum circuit learning algorithm \cite{huang2024learning}. The local inversion $V_i$ of $U$ on qubit-$i$ is an operator that disentangles the $i^\text{th}$-qubit from the rest of the circuit $U$
\begin{equation}
    V_i U = U' \otimes I_i.
\end{equation}
Note that in general, the remaining unitary $U'$ is different than $U$. Therefore we can not straightforwardly concatenate these inversions to reconstruct $U$. Instead, consider adding an ancilla qubit, and performing a 2-qubit SWAP before acting with the local inversion
\begin{equation}
    U V_1 S_1 = U' \otimes S_1,
\end{equation}
where $S_i$ is the 2-qubit SWAP between the original qubit-$i$ and the ancilla qubit-$i$. Then applying the inverse of the inversion $V_1^\dagger$ we obtain
\begin{equation}
    \label{eq:single-inversion}
    U V_1 S_1 V_1^\dagger = S_1 U.
\end{equation}
The circuit diagram illustrating this relation is given in figure \ref{fig:local-inversion}. If we add an ancilla qubit for each of our original qubits and act on with $V_i S_i V_i^\dagger$ for each qubit-$i$ from the right on equation \eqref{eq:single-inversion} we obtain
\begin{equation}
    (U \otimes I)\hat{U} = S (U \otimes I),  
\end{equation}
where the learned circuit is given by $\hat{U} = \prod_{i=1}^{n} V_i S_i V_i^\dagger$, and the $S = \prod_{i=1}^{n} S_i$ is the global SWAP gate between the ancilla qubits and the register. Inverting this equation, the learned circuit is given by
\begin{equation}
    \hat{U} = S(U \otimes U^\dagger).
\end{equation}
We can implement the original unitary by acting on the initial configuration $\rho \otimes \left|0^n\right>\left<0^n\right|$ with the learned circuit, and tracing out the ancilla qubits
\begin{equation}
    U \rho U^\dagger = \tr_{\text{Ancilla}} \left[S\hat{U}\left(\rho \otimes \left|0^n\right>\left<0^n\right| \right) (S\hat{U})^\dagger \right].
\end{equation}
Note that the learned circuit $\hat{U}$ is shallow because each $W_i = V_i S_i V_i^\dagger$ is local and is supported on the light-cone of qubit-$i$ and the ancilla-$i$. Therefore we can stack the operators $W_i$ that do not overlap simultaneously to construct a constant depth circuit.

Since we can construct a shallow circuit by the local inversions of the unitary, the problem of learning a shallow circuit is reduced to learning local inversions. The classical shadow of $U$ over $N$-samples is
\begin{equation}
    \cT_U(N) = \left\{\left|\psi_l\right> = \bigotimes_{i=1}^{n} \left|\psi_{l,i}\right>, \left|\phi_l\right> = \bigotimes_{i=1}^{n} \left|\phi_{l,i}\right>\right\}_{l=1}^{N},
\end{equation}
where each data sample specifies a random $n$-qubit product state $\left|\psi_l\right>$ and a randomized Pauli measurement outcome $\left|\phi_l\right>$ on the output state $U\left|\psi_l\right>$, where the single qubit states are stabilizer states. The algorithm utilizes a randomized measurement dataset $\cT_U(N)$. 

The central result of \cite{huang2024learning} is
\begin{theorem5}
    \label{thm:5}
    Given a failure probability $\delta$, an approximation error $\epsilon$, and an unknown $n$-qubit unitary $U$ generated by a constant-depth circuit over any two-qubit gates in a finite gate set of a constant size, then the algorithm learns an exact description $\cE = \cU$ with probability $1-\delta$, using $N=\cO(\log(n/\delta))$ samples and $t=\cO(\poly(n)\log(1/\delta))$ time.
\end{theorem5}
The key ideas for the proof require the following lemmas: \begin{lemma10}
    \label{lem:10}
    Given a failure probability $\delta$, an approximation error $\epsilon$, an unknown $n$-qubit observable $O$ with $||O||_\infty \leq 1$, that acts on an unknown set of $k$ qubits, and a dataset $\cT_O(N)=\{\left|\psi_l\right> = \bigotimes_{i=1}^{n} \left|\psi_{l,i}\right>, v_l\}_{l=1}^{N}$ where $\left|\psi_{l,i}\right>$ is uniformly sampled from stab${}_1$ and $v_l$ is a random variable with $\mathbb{E}[v_l]=\left<\psi_l\right|O\left|\psi_l\right>$, $|v_l|=\cO(1)$. Given a dataset of size
    \begin{equation*}
        N = 2^{\cO(k)}\log(n/\delta)/\epsilon^2,
    \end{equation*}
    we can learn an observable $\hat{O}$ such that $||\hat{O}-O||_\infty\leq\epsilon$ and $\text{supp}(\hat{O})\subseteq\text{supp}(O)$ with probability at least $1-\delta$.
\end{lemma10}

\begin{lemma13}
    \label{lem:13}
    Given $3n$ $n$-qubit operators $\hat{O}_{i,P}$ where $i\in\{1,\dots,n\}$ and $P\in\{X, Y, Z\}$ such that for any qubit $i$, $\left| \bigcup_P \text{supp}(\hat{O}_{i,P})\right| = \cO(1)$ and there is only a constant number of qubits $j$ with
    \begin{equation*}
        \left(\bigcup_P\text{supp}(\hat{O}_{i,P})\right)\cap\left(\bigcup_P\text{supp}(\hat{O}_{j,P})\right)\neq\emptyset,
    \end{equation*}
    there exists a sewing order for $U_{\text{sew}}(\left\{O_{i,P}\right\}_{i,P})$ that can be implemented by a constant depth circuit.
\end{lemma13}

\begin{lemma9}
    \label{lem:9}
    Given an $n$-qubit unitary $U$, and  $3n$ $n$-qubit operators $O_{i,P}$ where $i\in\{1,\dots,n\}$ and $P\in\{X, Y, Z\}$. Assume $O_{i,P}$ is an $\epsilon_{i,P}$-approximate Heisenberg-evolved Pauli observable $P$ on qubit-$i$ under $U$. Let $U_{\text{sew}}=U_{\text{sew}}(\left\{O_{i,P}\right\}_{i,P})$. Then
    \begin{equation*}
    \cD_\diamond(\cU_{\text{sew}},\cU\otimes\cU^\dagger)=\frac{1}{2}\left|\left|\cU_{\text{sew}}-\cU\otimes\cU^\dagger\right|\right|_\diamond\leq \sum_i \sum_P \epsilon_{i,P}.
    \end{equation*}
\end{lemma9}
The diamond-distance $\cD_\diamond(\cE_1,\cE_2) = \left|\left|\cE_1 - \cE_2\right|\right|_\diamond$ of two completely positive, trace non-increasing maps $\cE_1$ and $\cE_2$ is given by the diamond-norm which is defined as \cite{kitaev1997quantum}
\begin{equation}
    \left|\left|\cE_1 - \cE_2\right|\right|_\diamond = \max_{\rho} \left|\left|(\cE_1\otimes I_n)\rho - (\cE_2\otimes I_n)\rho\right|\right|_1
\end{equation}
where $\left|\left| \cdot \right|\right|_1$ is the trace-norm.

In essence, the question of learning shallow circuits is divided into two separate problems. Learning the local inversions, and sewing them back together. Instead of learning the whole unitary $U$, it's much easier to learn its local inversions $\{V_i\}$. If one learns the Heisenberg-evolved Pauli operators $U^\dagger P_i U$ for all $P\in\{X,Y,Z\}$, then local inversions can be found that reverse the Heisenberg evolution, hence minimizing
\begin{equation}
    \label{eq:brute}
    \sum_{P\in\{X,Y,Z\}}\left|\left|V_i^\dagger U^\dagger P_i U V_i - P_i \right|\right|_\infty.
\end{equation}
Having learned the local inversions, one must ``sew'' them back using the ancilla qubits and SWAP gates outlined earlier to reconstruct the original unitary. To that end Lemma 10 is used to learn approximate Heisenberg-evolved Pauli observables that will be used in the brute force search for local inversions when minimizing \eqref{eq:brute}, Lemma 13 shows that the learned Heisenberg-evolved Pauli observables can be sewn to form a constant depth circuit, and Lemma 9 provides the rigorous performance guarantee that the sewn circuit indeed approximates $U$.

To apply the LIL algorithm, the underlying circuit $U$ does not have to be shallow, it can be a deep circuit. The algorithm will still yield a learned circuit $\cE$, but the learned circuit might not be close to the $n$-qubit channel $\cC(\rho) = U \rho U^\dagger$ that we would like to implement. An efficient verification algorithm is presented in \cite{huang2024learning}
\begin{theorem10}
    \label{thm:10}
    Given a failure probability $\delta$, an approximation error $\epsilon$, a learned constant depth $2n$-qubit circuit $\hat{V}$, the associated CPTP map $\hat{\cE} = \cE_{\leq n}^{\hat{V}}$ and an unknown CPTP map $\cC$, a randomized measurement dataset $\cT_{\cC}(N)$ of size $N=\cO(n^2\log(n/\delta)/\epsilon^2)$. The verification algorithm outputs PASS with probability $\geq 1-\delta$ if $\cD_\text{ave}(\hat{\cE},\cC)\leq\epsilon/12n$ and $\left|\left|\cC^\dagger\cC-\cI\right|\right|_\diamond\leq \epsilon/12n$ and outputs FAIL with probability $\geq 1-\delta$ if $\cD_\text{ave}(\hat{\cE},\cC)>\epsilon$.
\end{theorem10}
The learned circuit $\cE$ passes the test if it's close to the channel $\cC$ in average case distance and the channel is unitary. This is the exact reason why we can bootstrap the shallow quantum circuit learning algorithm to compress deep quantum circuits. Given a sub-circuit, we can learn a shallow circuit and verify if it reproduces the appropriate channel. If it passes the verification, we replace it with the learned shallow circuit which leads to a compression of the circuit depth. If it fails the verification, we cut into smaller sub-circuits, until all the circuits pass the verification algorithm. 

Since we will be recursively applying the LIL algorithm, we are only interested in the case where the shallow representation of each sub-unitary exactly reproduces it, which restricts us to learn over a finite gate set. Approximate learning in diamond distance is also studied for continuous gate sets e.g. $SO(4)$, but this wouldn't be appropriate in circuit optimization since the combination of epsilon-close in diamond distance circuits is not epsilon-close in diamond distance necessarily once enough compositions have been performed. Thus, in this scenario, only a polynomial number of compressions are allowed to occur to retain a controlled error rate. When learning over finite gate sets, the circuit $U$ can be learned to zero error with high probability $1-\delta$ from $N=\cO(\log(n/\delta))$ samples in $t=\cO(\poly(n)\log(1/\delta))$ time.

The authors have shown that logarithmic depth circuits require exponentially many queries to the unitary to learn within a small diamond distance. Our work circumvents this problem by enabling shallow circuit learning in higher-depth circuits via the recursive application of their algorithm.

\begin{figure*}
    \includegraphics[width=\textwidth]{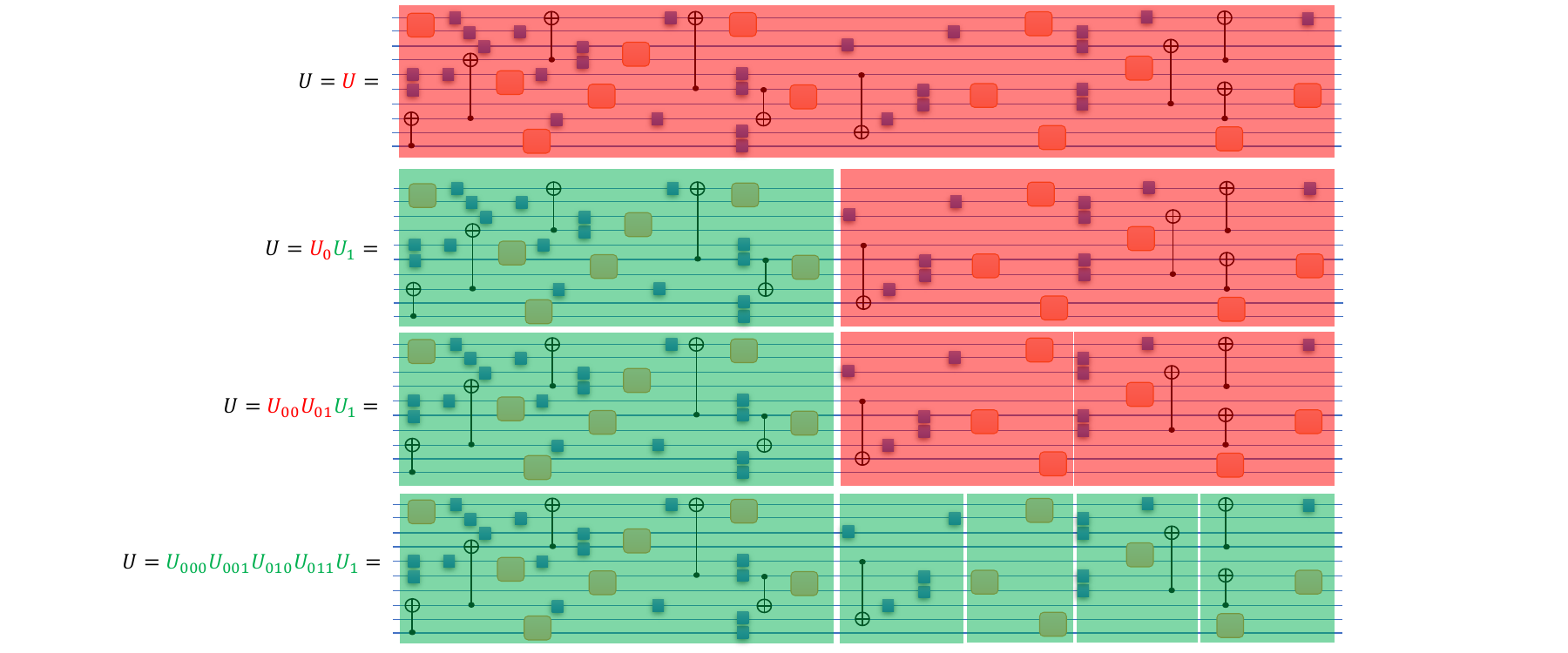}
    \caption{Binary search for shallow representations. The green sub-circuits have passed the LIL algorithm, and the red sub-circuits have failed. We start with the original circuit and apply the LIL algorithm. Then we cut the circuit in half and applied it again. We iterate this process until all the sub-circuits have been replaced with their shallow representations.}
    \label{fig:binary-search}
\end{figure*}

\section{Binary search for shallow representations}
\label{sec:compression}

We now show how the LIL algorithm can be iterated to optimize quantum circuit depth. Given a quantum circuit, certain regions might have been inefficiently realized, i.e. the unitary that a sub-region implements might have a shorter depth representation. In particular, the unitary that the sub-circuit implements might have a shallow representation. If the shallow representation exactly reproduces the original sub-unitary, replacing the sub-circuit with its shallow representation compresses that portion of the circuit. Therefore given a circuit $U$, we can recursively apply LIL to find constant depth representations of sub-circuits, which will lead to an overall compression in circuit depth. The best-case scenario is that the initial circuit itself passes the LIL algorithm, which results in a total compression of the circuit. The worst-case scenario is that the recursive search halts only when all the sub-circuits are comprised of a single gate, in which case they are trivially shallow. Therefore, depending on the given circuit, our algorithm can realize any compression rate between 0 and 1. Due to the binary search structure of the algorithm, it has time complexity $t = \cO(\log d)$. However, since the LIL of each sub-circuit in the binary search tree is independent of one another, the compression algorithm is massively parallelizable. Unfortunately, a test of our compression algorithm can not be carried out at the time of writing, since the numerical implementation of the shallow circuit learning algorithm does not exist. The pseudocode \eqref{alg:compress} shows how our compression algorithm works, using the local inversion learning and the verification algorithm of \cite{huang2024learning}.

\begin{algorithm}
    \caption{Shallow compression algorithm}
    \label{alg:compress}
    \begin{algorithmic}[1]
        \Procedure{lil}{$U$}
            \State $\{V_i\} \gets$ brute force search for local inversions of $U$
            \State $\hat{U} \gets$ sew local inversions $\{V_i\}$ of $U$
            \State return $\hat{U}$
        \EndProcedure
        \State
        \Procedure{ver}{$U$, $\hat{U}$} 
            \State $\cC^{U}, \cC^{\hat{U}}$ $\gets$ associated CPTP maps of $U$ and $\hat{U}$ 
            \If{$\cD_\text{ave}(\cC^{\hat{U}}, \cC^{U}) \leq \frac{\epsilon}{12}$ and $\left|\left|\cC^{U\dagger}\cC^{U} - \cI \right|\right|_\diamond \leq \frac{\epsilon}{12}$}
                \State return TRUE
            \ElsIf{$\cD_\text{ave}(\cC^{\hat{U}}, \cC^{U}) > \epsilon$}
                \State return FALSE
            \EndIf
        \EndProcedure
        \State
        \Procedure{cut}{$g$}
            \State $g_L$ and $g_R$ $\gets$ left and right half of the circuit $g$
            \State return $\{g_L,g_R\}$
        \EndProcedure
        \State
        \State $S \gets \{ U\}$
        \State $\hat{U} \gets \Call{lil}{U}$
        \State compressed $\gets \Call{ver}{\hat{U},U}$
        \While{$\neg$ compressed}
            \State compressed $\gets$ True
            \For{$g\in S$}
                \State $\hat{g} \gets \Call{lil}{g}$
                \If{$\Call{ver}{\hat{g},g}$}
                    \State $g \gets \hat{g}$
                \Else
                    \State compressed $\gets$ False
                    \State $g \gets \Call{cut}{g}$
                \EndIf
            \EndFor
        \EndWhile
        \State $\hat{U} \gets \prod_{g\in S} g$
        \State return $\hat{U}$
    \end{algorithmic}
\end{algorithm}

The recursive search and replacement of shallow components do not lead to a build-up of errors in the case of a finite gate set because the learning algorithm is exact. This compression would not be possible if we had used a continuous gate set, in which case the learned shallow representation simulates the original unitary only up to a small diamond distance, and by replacing sub-circuits with $\epsilon$-approximate shallow representations, it is not guaranteed that the compressed circuit is $\epsilon$-close to the original unitary unless the number of compressions allowed was restricted in an appropriate way.

As we cut the circuit in half recursively, the sub-circuits form a binary tree structure. We can label the circuits with binary strings. Since the total depth of the circuit is $d$, each sub-circuit represented by a binary string $l$ has depth $d(l) = 2^{-|l|} d$, where $|\cdot|$ is the length of a binary string. Hence, if a sub-circuit $U_l$ is replaced with its shallow representation, its depth compresses from $\cO(2^{-|l|} d) \to \cO(1)$. Therefore, if the set of binary strings that label the sub-circuits amenable to shallow representation is $S$, then the total compression rate achieved is given by
\begin{equation}
    R = 1 - \sum_{l\in S} 2^{-|l|}.
\end{equation}
If the LIL only succeeds for single-gate sub-circuits, the compression rate is given by
\begin{equation}
    R_{\text{single-gate}} = 1 - 2^{-\log(d)} d = 0.
\end{equation}
Therefore trivial shallow-representations do not contribute to circuit compression.

\section{Discussion}
\label{sec:discussion}

In this work, we have shown that the shallow quantum circuit learning algorithm can be bootstrapped to compress arbitrary quantum circuits. The recursive structure bypasses the problems faced in generalizing local inversion learning to higher-depth circuits. We have only discussed recursive half-cuts, but varying cut ratios can be tested. A further question of interest is if there is an optimal cut for circuit compression. Classical machine learning techniques can be utilized to learn the optimal cut ratio by training the compression algorithm on random unitary circuits. A priori, there is no reason to expect a universal optimal cut value for all unitaries, but the existence of an optimal cut for different families of unitaries might be an important tool in the understanding and classification of quantum algorithms. Rather than cutting the circuit into two pieces iteratively, one can also consider optimizing the bulk of the circuit. We can decompose the circuit into 3-parts by selecting two time slices $t_i$ and $t_f$
\begin{equation}
    U = U(t_f) U(t_i;t_f) U(t_i).
\end{equation}
By looping over $t_i$ and $t_f$, we can compress $U(t_i;t_f)$ by finding a shallow representation for the highest depth $t_f-t_i$. One can also directly partition the circuit into $n$ sub-circuits, each with depth $\theta_i$ for $i \in \{1,\dots,n\}$. Then by searching for a shallow representation for each sub-circuit, one can achieve a compression rate $R_{\theta}(U)$. More generally, rather than having a fixed cut ratio for all legs of the binary search tree, one can start from a random probability distribution on cuts $P(\theta|x)$, where $\theta$ is the cut ratio at decision point $x$. An optimization agent applies the shallow learning algorithm along the sub-circuits, and updates the cut distribution based on the success rate achieved at each leg. Shallow learning at earlier stages of the binary tree is more rewarding because the sub-circuit depth decreases exponentially as one goes down the binary tree. We hope to report on numerical simulations of practical circuit compression and learning optimal cuts in future work.

\section*{Acknowledgements}
We thank Jim Halverson, Stephen Jordan, Isaac Kim, Joydeep Naskar, Fabian Ruehle, and Henry Yuen for useful discussions. N.B. is supported by the DOE Office of Science ASCR, in particular under the grant Novel Quantum Algorithms from Fast Classical Transforms. G.S. is supported by the Graduate Assistantship from the Department of Physics, Northeastern University.

\bibliographystyle{IEEEtran}
\bibliography{ref.bib} 

\end{document}